\documentclass[twocolumn]{aastex631}

\usepackage{textcomp}

\usepackage{multirow}

\newcommand{\dmunits}{\ensuremath{{\,{\rm pc}\,{\rm cm}^{-3}}}\xspace}

\newcommand*\subtxt[1]{_{\textnormal{#1}}}
\DeclareRobustCommand\_{\ifmmode\expandafter\subtxt\else\textunderscore\fi}

\usepackage{color}
\usepackage{graphicx, epstopdf} 
\usepackage{gensymb}
\usepackage{natbib}
\usepackage{amsmath}
\usepackage{hyperref}
\usepackage{amsfonts}
\usepackage{amsthm}
\usepackage{xspace}
\usepackage{multirow}
\usepackage{url}
\usepackage{placeins}

\received{February 17 2026}
\accepted{March 12 2026}
\submitjournal{ApJ}

\shorttitle{FAST-GBT Discovery \& Timing Analysis}
\shortauthors{Blackmon et al.}

\begin{document}

\title{Discovery and Timing Follow-Up of Two FAST-Discovered Pulsars from the FAST CRAFTS Survey}

\author{Victoria A. Blackmon}
\affiliation{West Virginia University, 1500 University Ave, Morgantown, WV, 26506}
\affiliation{Center for Gravitational Waves and Cosmology, West Virginia University, Chestnut Ridge Research Building, Morgantown, WV 26506, USA}

\author{Maura A. McLaughlin}
\affiliation{West Virginia University, 1500 University Ave, Morgantown, WV, 26506}
\affiliation{Center for Gravitational Waves and Cosmology, West Virginia University, Chestnut Ridge Research Building, Morgantown, WV 26506, USA}

\author{De Zhao}
\affiliation{National Astronomical Observatories, Chinese Academy of Sciences, Beijing 100012, China}

\author{Jianping Yuan}
\affiliation{National Astronomical Observatories, Chinese Academy of Sciences, Beijing 100012, China}

\author{Qingdong Wu}
\affiliation{National Astronomical Observatories, Chinese Academy of Sciences, Beijing 100012, China}

\author{Chen-Chen Miao}
\affiliation{National Astronomical Observatories, Chinese Academy of Sciences, Beijing 100012, China}

\affiliation{School of Astronomy and Space Science, University of Chinese Academy of Sciences, Beijing 100049, China}

\author{Meng-Yao Xue}
\affiliation{National Astronomical Observatories, Chinese Academy of Sciences, Beijing 100012, China}

\author{Di Li}
\affiliation{New Cornerstone Science Laboratory, Department of Astronomy, Tsinghua University, Beijing 100084, China}
\affiliation{National Astronomical Observatories, Chinese Academy of Sciences, Beijing 100012, China}

\author{Wei-Wei Zhu}
\affiliation{National Astronomical Observatories, Chinese Academy of Sciences, Beijing 100012, China}
\affiliation{Institute for Frontiers in Astronomy and Astrophysics, Beijing Normal University, Beijing 102206, China}

\begin{abstract}

We present the results of Green Bank Telescope (GBT) observations of two pulsars discovered with the Five-hundred-meter Aperture Spherical Radio Telescope (FAST) during the 19-beam Commensal Radio Astronomy FasT Survey (CRAFTS). We highlight the first timing solutions, pulse profiles, flux densities, and polarization measurements at 820 MHz for PSR~J0535--0231, with a spin period of 415 ms, and  PSR~J1816--0518, with a spin period of 1.93 s, from a year-long follow-up campaign. PSR~J0535--0231 appears to be partially recycled, but isolated, and likely belongs to the class of disrupted recycled pulsars (DRPs). We find that the two widely used electron density models, NE2001 and YMW16,  both fall short of accurately modeling the line-of-sight  to PSR~J0535--0231, as the maximum dispersion measure (DM) predicted by both models is lower than the pulsar's DM  of 118.1 \dmunits. Finally, we place both pulsar discoveries in the context of other FAST pulsars discovered in the CRAFTS survey and of the currently known pulsar population, in general, and discuss ways in which future FAST discoveries of faint, distant pulsars might facilitate the development of improved versions of the aforementioned electron density models in certain regions of our Galaxy.

\end{abstract}


\section{Introduction}

Rapidly rotating, highly magnetized neutron stars, often observable as radio pulsars, emerge from core-collapse supernova explosions of   stars with masses between approximately 8 and 25 M$_\odot$.
Upon formation, most radio pulsars exhibit rotational periods on the order of tens to hundreds of milliseconds and magnetic field strengths ranging  between 10$^{11}$ and 10$^{13}$ G \citep{Igoshiev_2021}.
These pulsars generate radio emission for up to a  billion years as their spin periods gradually increase as a result of energy loss due to magnetic dipole radiation \citep{OG_1969}. In general, these ``non-recycled'' pulsars provide valuable insights into neutron star evolution, the pulsar population, emission mechanisms, Galactic dynamics, and the interstellar medium \citep{Lorimer_2004}.

Alternatively, recycled pulsars are those that have been spun-up through the accretion of matter from an evolving binary companion. Those with higher-mass companions may be partially recycled to periods of tens of milliseconds and magnetic field strengths of $10^9$-$10^{11}$~Gauss before their companions undergo a supernova explosion. These objects are detectable as double neutron star (DNS) binaries or disrupted recycled pulsars (DRPs), depending on whether the binary system remains intact after the supernova event. Those with lower-mass companions will be fully recycled to spin periods on the order of milliseconds and magnetic field strengths between 10$^{8}$ and 10$^{10}$ G. Millisecond pulsars are very stable rotators, some of which allow for timing accuracy down to sub-microsecond level precision. They serve as effective tools for fundamental physics, such as neutron star mass measurements or the search for and detection of gravitational waves emanating from distant sources beyond our Galaxy. For these reasons, the discovery of pulsars is an ongoing effort that is vital to the  advancement of astrophysics.

Since their initial discovery \citep{Hewish_1968}, the majority of presently known radio pulsars have been found through un-targeted sky surveys in which all or part of a given radio telescope's available sky is scanned for pulsar-like signals. 
In 2019, a drift-scan survey with the Five-hundred-meter Aperture Spherical Telescope (FAST) began, aiming to search the entire observable sky for pulsars and radio transients and carry out spectral and imaging studies on galaxies and neutral hydrogen \citep{Li_18}. This Commensal Radio Astronomy FAST Survey, or CRAFTS, predicted the discovery of more than one thousand pulsars using the FAST 19-beam L-band receiver \citep{dunning_17}.

Thus far, over two hundred pulsars have been discovered by FAST in the CRAFTS survey \citep{FAST-CRAFTS}, including at least 74 millisecond pulsars, approximately 141 non-recycled pulsars and about seven confirmed rotating radio transients (RRATs). Given the large number of discoveries in this survey, follow-up observations of newly-discovered pulsars using other large radio telescopes  are necessary to confirm and characterize these pulsars and obtain timing solutions.

In this paper, we describe our efforts to produce combined FAST and Green Bank Telescope (GBT) timing solutions spanning several years for two CRAFTS discoveries --
PSRs~J0535--0231 and J1816--0518 --
and present polarization profiles for both sources, based on FAST and GBT data at each observing frequency.

The subsequent sections of this paper are organized as follows. In section 2 we provide the observation details used in our GBT follow-up campaign and describe the process by which times-of-arrival (TOAs) were generated. In section 3, we introduce PSRs J0535--0231 and J1816--0518, outlining additional data reduction steps and our method of producing polarization profiles. In section 4, we discuss timing solutions and profile evolution and place these pulsars in the context of the broader pulsar population. We also discuss the shortcomings of NE2001 and YMW16 regarding the dispersion measure (DM) distance modeling for PSR~J0535--0231 as well as ongoing efforts in pulsar searching with FAST and future plans for follow-up work.

\section{Observations \& Analysis} 

Using the 105-m GBT in Green Bank, West Virginia we  carried out an observational campaign focused on  four FAST-discovered pulsars. These pulsars include partially-recycled pulsar, PSR~J0535--0231, with a spin period of 415 ms and DM of 118.1 \dmunits and non-recycled PSR~J1816--0518, with a spin period of 1.93 s and DM of 137.7 \dmunits.
We also observed the 30.72-ms period double neutron star (DNS) binary PSR J0631+4147 and black widow (BW) PSR J1720--0534, with spin period of 3.26 ms.

We initially proposed for a year-long campaign of approximately monthly half-hour follow-up observations of PSRs~J0535--0231 and J0631+4147, along with other pulsars we were ultimately unable to detect, aiming to produce initial timing solutions for each.
Each GBT observation was coherently dedispersed with full-Stokes polarization, and run in VEGAS Pulsar Mode (VPM) using the 800-MHz Prime Focus 1 Receiver at 820-MHz center-frequency with a 200-MHz bandwidth, a frequency resolution of 0.78 MHz and  a sampling time of 10.24 µs. Additionally, we also completed a single observation of PSR~J0535--0231 at L-Band (1500-MHz center frequency) with an 800-MHz bandwidth under the same observing setup.  Approximately two months after the start of our observations, we added PSR~J1816--0518 to our campaign, aiming to produce enough TOAs for an initial timing solution. The initial six 820-MHz observations of this pulsar were 40--60 minutes each. The latter three observations were  10--20 minutes, sufficient to obtain a high signal-to-noise detection. In two observations for which this pulsar was undetectable, we first performed radio frequency interference (RFI) removal in our data reduction routine but ultimately attribute these non-detections to the reduction of the overall signal-to-noise due to narrowband RFI. The observing campaign began in May 2022 and we observed each pulsar with a monthly cadence. In the final four months of the observing campaign, under the same observing specifics at 820 MHz, we added monthly observations of PSR J1720--0534, but did not obtain enough data to produce an initial timing solution, hence its exclusion from further analysis. The number of detections per source and total accumulated observation times are summarized in Table~\ref{tab:observing}.
\begin{deluxetable}{ccccc}
\tablewidth{\textwidth} 
\tablecaption{GBT observation details for the year-long follow-up campaign where the number of observations, N$\subtxt{obs}$, and the total integration time in hours, T$\subtxt{obs}$, are given.}
\tablehead{& \colhead{PSR} & \colhead{Detections} &
 \colhead{N$\subtxt{obs}$} & \colhead{T$\subtxt{obs}$ (h)}} 
\startdata
\multirow{5}*{} 
& J0535--0231 & 12 & 12 & 2.97  \\
& J1816--0518 & 9 & 10 & 4.44 \\
& J0631+4147 & 8 & 12 & 2.21 \\
& J1720--0534 & 4 & 4 & 0.79 \\
\hline
\enddata

\end{deluxetable}
\label{tab:observing}

To produce GBT TOAs, we began by folding the raw data using \texttt{DSPSR} \citep{2011PASA...28....1V} from the \texttt{PSRCHIVE}\footnote{\url{https://psrchive.sourceforge.net/}} library \citep{2004PASA...21..302H}. RFI mitigation required use of the software \texttt{pazi} to excise RFI in frequency and time. We then used \texttt{pat} to calculate TOAs by cross-correlating folded pulse profiles from each epoch with a high S/N composite template, produced by summing average profiles across all observations, and ultimately obtained {one TOA per epoch for PSR~J0535--0231 and PSR~J1816--0518.

FAST employs a 19-beam receiver with an effective bandwidth of 400 MHz (1050$-$1450 MHz) and a sampling time of 49.152 $\mu$s \citep{2020RAA....20...64J}. We utilized 40 s of noise-diode data collected prior to each observation for polarization calibration. The FAST data were first folded with \texttt{DSPSR} using the initial spin period and dispersion measure obtained from \texttt{PRESTO}\footnote{\url{https://github.com/scottransom/presto}}. This produced integrated profiles with 512 phase bins per epoch. Additionally, RFI was excised using \texttt{PSRCHIVE}. After aligning and summing all epochs, we utilized \texttt{paas} to generate a standard template and generated TOAs with \texttt{pat}.

We calculated timing solutions using the pulsar software \texttt{TEMPO2}\footnote{\url{https://www.atnf.csiro.au/research/pulsar/tempo2}},  which performs least-squares fits to minimize the reduced chi-squared of the timing residuals. We applied the built-in EFAC to adjust the TOA uncertainties such that our resulting $\chi^2$ was approximately 1. TOAs were phase-connected across the span of our data, and we fit for astrometric, spin, and interstellar medium (ISM) parameters, namely, right ascension ($\alpha$), declination ($\delta$), spin frequency ($\nu$), spin-down rate ($\dot\nu$), and dispersion measure (DM). We also fit a JUMP between the GBT and FAST observations.

For PSR~J0535--0231, we first calculated FAST TOAs with  4 subbands and 4-second subintegrations, and fit those to the full timing model, including DM. In order to increase the TOA signal-to-noise, we then calculated only one TOA per epoch for all 23 FAST epochs and fit them with the GBT TOAs, keeping the DM fixed to avoid degeneracy with the JUMP parameter.

Due to the small number of FAST observations for PSR~J1816$-$0518,  we obtained TOAs in  4 subbands with approximately 4-minute subintegration lengths.  
In this paper, we present timing solutions for two pulsars -- PSRs~J0535--0231 and J1816--0518 -- that span just over a year-long baseline including data from both FAST and the GBT. These timing solutions are shown in Table~\ref{tab:timing}.  The DM and error reported for PSR~J0535--0231 are derived from the fit to the subbanded FAST data.

We  also list timing-derived inferred surface magnetic field strength ($B$),  characteristic age ($\tau$) and  spin-down luminosity ($\dot E$) for each pulsar. Measurements of these parameters allow us to place the pulsars in the context of the general pulsar population, as shown in the period--period derivative ($P$--$\dot P$) diagram in section 4. We do not present timing solutions for BW pulsar J1720--0534 and DNS pulsar J0631+4147. The timing solution, which includes GBT data, for PSR~J1720--0534 is presented in \cite{Miao_2023} and that of PSR~J0631+4147 is pending publication by the same authors. However, we include them in this population figure for completeness.

\section{Timing Solutions \& Profiles}
In the last four months of the campaign, we completed full-polarization calibration scans on-source for each pulsar and minute-long on- and off-source flux calibration scans of the nearby, bright radio galaxy 3C123. Using \texttt{PSRCHIVE}, we utilized the function \texttt{pac} along with these scans to produce fully calibrated profiles for PSRs~J0535--0231 and J1816--0518 at each epoch. We then combined profiles at each epoch to create composite profiles at a given observing frequency using \texttt{psradd}.

Following this process, we  performed rotation measure (RM) fits, using the \texttt{rmfit} function, to each profile to correct for rotation of the linear polarization position angle due to the Faraday effect. These RM measurements were then applied to improve the accuracy of each profile. For PSRs J0535--0231 and J1816--0518 we find  RMs of 77 $\pm$ 11 rad m$^{-2}$ and --21 $\pm$ 7 rad m$^{-2}$, respectively.
The GBT composite profiles for PSR J0535--0231 at 820 MHz and L-Band are shown in Figure~\ref{fig:image1} (a,c), supplemented by the FAST L-Band profile in Figure~\ref{fig:image1} (b). We also include the profiles for PSR J1816--0518 from the 820-MHz GBT observations and the L-Band FAST observations in Figure~\ref{fig:image2}. The timing residuals for both pulsars are given in Figure~\ref{fig:image3}, where blue and green residuals correspond to GBT 820-MHz and FAST L-Band data, respectively.

\begin{deluxetable*}{cccc}
\tablewidth{\textwidth} 
\label{tab:FAST_canon}
\tablecaption{TEMPO2 timing parameters for FAST-discovered PSRs J0535--0231 and J1816--0518}
\tablehead{& \colhead{PSR} & \colhead{J0535--0231} & \colhead{J1816--0518}} 

\startdata
\multirow{6}*{Timing Data} & Data span (yr) & 1.8 & 1.4 \\
& Start epoch (MJD) & 59347 & 59399 \\
& End epoch (MJD) & 60002 & 59892 \\
& Timing epoch (MJD) & 59664 & 59635 \\
& Number of TOAs & 34 & 28 \\
\hline
\multirow{5}*{Measured Parameters} & 
Right Ascension, $\alpha$ (J2000) & $05^{\rm h}\, 35^{\rm m}\, 34\, \fs6341(4)$ & $18^{\rm h}\, 16^{\rm m}\, 45\fs 926(7)$ \\ 
& Declination, $\delta$ (J2000) & $-02\arcdeg\, 31\arcmin\, 50\, \farcs196(14)$ & $-05\arcdeg\, 18\arcmin\, 02\, \farcs0(3)$  \\ 
& Spin frequency, $\nu$ (Hz) &  2.4095598132(3) & 0.5180551655(13)  \\
& Spin frequency derivative, $\dot{\nu}$ (Hz s$^{-1}$) & 
$-3.61$(3) $\times$ 10$^{-17}$ & 
$-3.1431$(15) $\times$ 10$^{-15}$\\
& Dispersion measure (pc cm$^{-3}$) & 118.1(9) & 137.7(4) \\ 
\hline
\multirow{2}*{Fitting Parameters} &
RMS post-fit residuals ($\mu$s) & 32.5 & 171.4\\
& EFAC & 1.1 & 1.11 \\
\hline
\multirow{10}*{Derived Parameters} & Spin period, $P$ (s) &  0.415013561 &  1.93029636 \\ 
&  Spin period derivative, $\dot{P}$ (s s$^{-1}$) &  6.217 $\times$ 10$^{-18}$ & 1.171 $\times$ 10$^{-14}$  \\
& Galactic Longitude, $\ell$ ($\degree$) & 206.366 & 24.231 \\
& Galactic Latitude, $b$  ($\degree$) & --18.004 & 5.246 \\
& DM Distance (kpc) -- NE2001 & 44.4 & 3.5 \\
& DM Distance (kpc) -- YMW16 & 25 & 4.3 \\
& Rotation Measure (rad m$^{-2}$) & 77 $\pm$ 11 & --21 $\pm$ 7\\
& Mean Flux Density (mJy) & 0.020 $\pm$ 0.003  & 0.0119 $\pm$ 0.0015 \\
& Spin-down luminosity, $\dot{E}$ (erg s$^{-1}$) & 3.5 $\times$ 10$^{30}$ & 6.5 $ \times$ 10$^{31}$ \\
& Inferred surface dipole magnetic field, $B_{\rm{surf}}$ (G) & 5.1 $\times$ 10$^{10}$ & 4.8 $\times$ 10$^{12}$ \\ 
& Characteristic age, $\tau_c$ (Myr) & 1047 & 2.6  \\
\hline
\enddata
\tablecomments{These timing solutions make use of the JPL DE440 solar system ephemeris and values in parentheses indicate TEMPO2-reported 1$\sigma$ uncertainties on the least-significant digits. Standard deviations are reported for the mean flux values.}
\end{deluxetable*}
\label{tab:timing}

\begin{figure*}
    \gridline{\fig{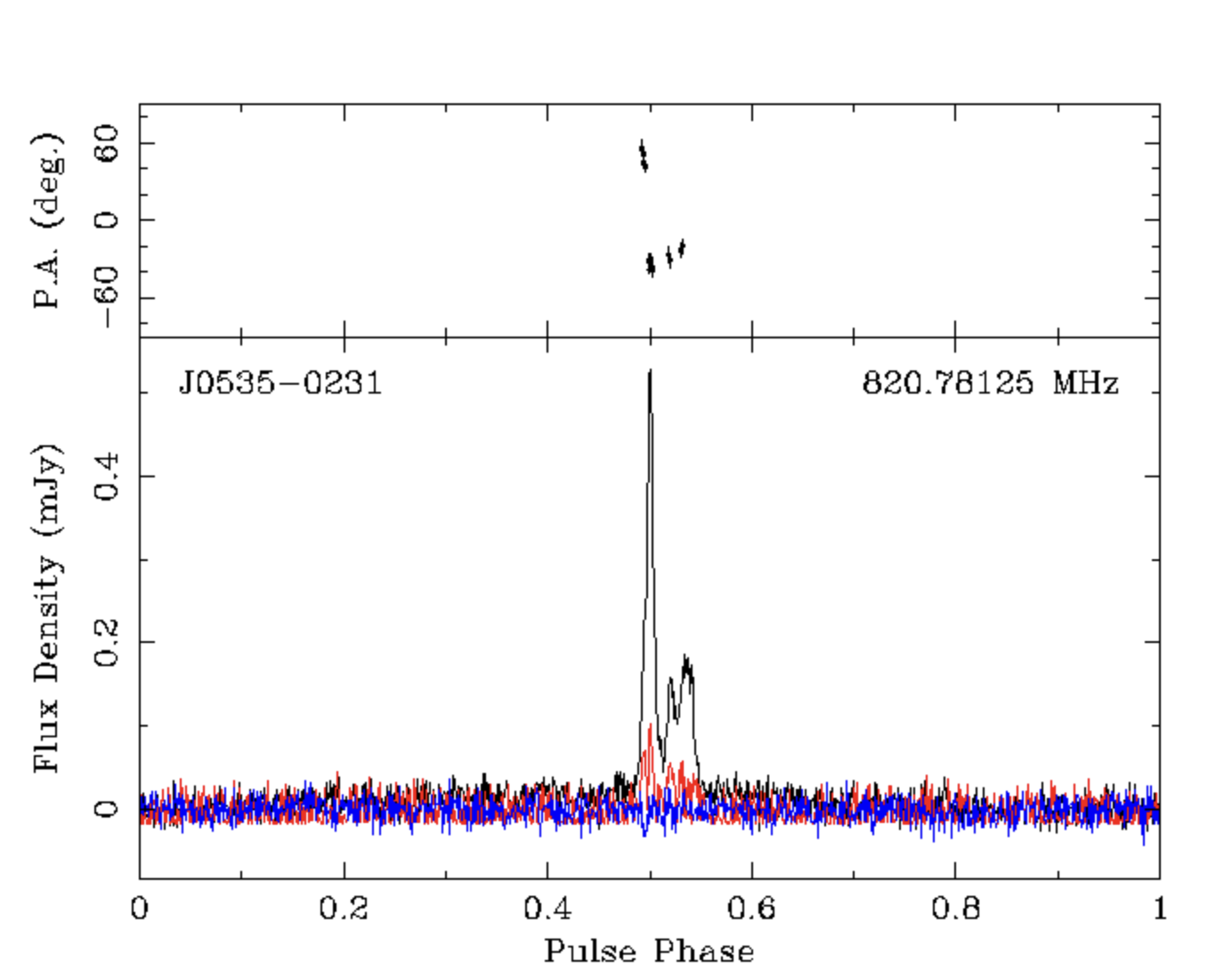}{0.49\textwidth}{(a) Flux and polarization calibrated GBT 820-MHz composite profile based on an  2.97 hours of data.}
              \fig{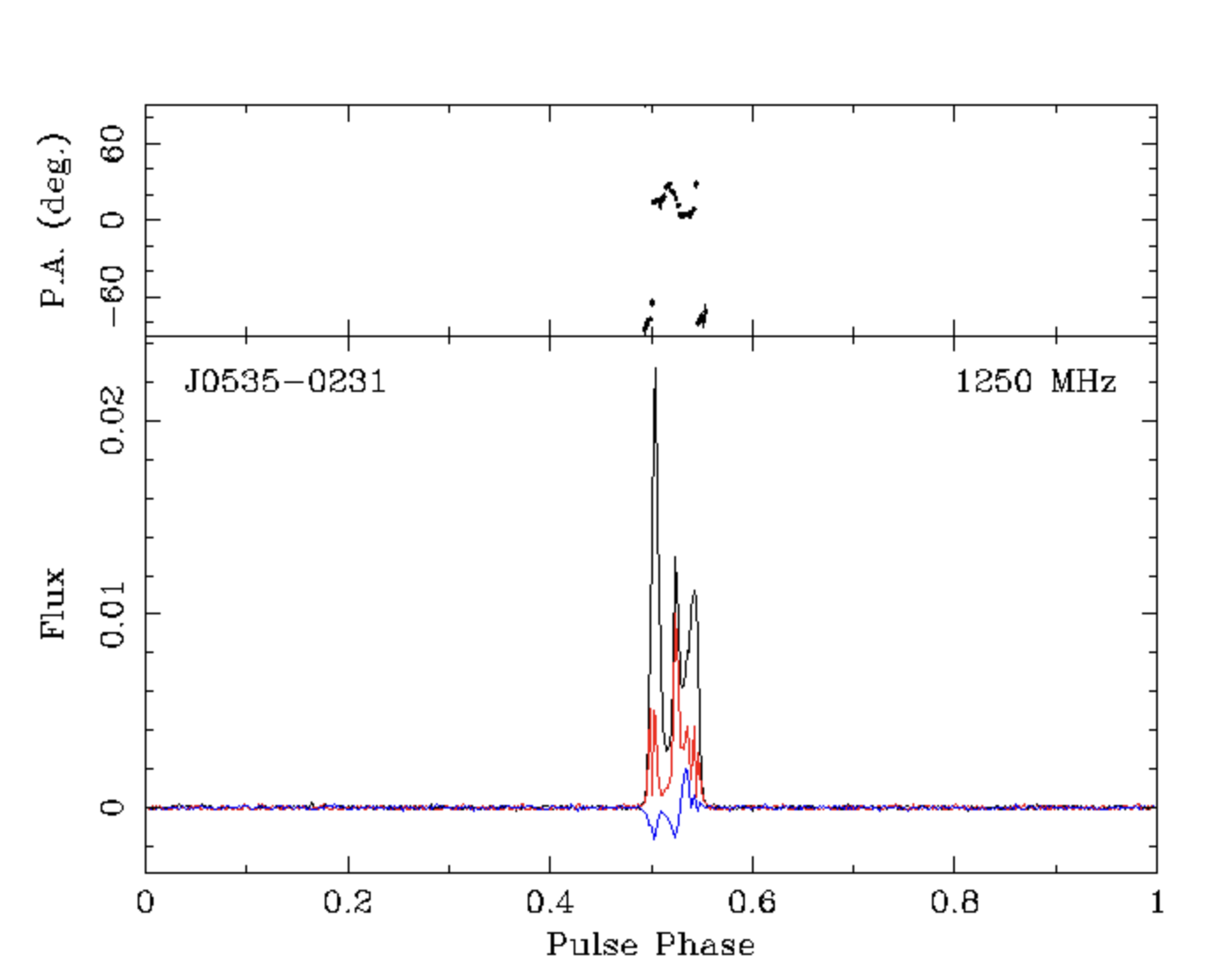}{0.49\textwidth}{(b)  FAST L-band composite profile (in arbitrary flux units) based on  1.37 hours of data. }
                 }

    \gridline{\fig{Updated_GBT_LBand.png}{0.55\textwidth}{(c)  GBT L-band total intensity profile (in arbitrary flux units) based on 0.12 hours of data. 
             }}
             
 \caption{Profiles of PSR~J0535--0231 at different observing frequencies. All profiles are shifted such that the peak flux component is centered at a phase of 0.5. (a)  is the profile based on GBT data at 820 MHz, (b) is the profile at L band (1250 MHz) based on the FAST data, and (c) is the profile of the single L-band (1500 MHz) GBT observation. Stokes parameters I, L, and V are represented by black, red, and blue curves, respectively. The top panel in each plot indicates the position angle of the linear polarization as the pulsar rotates.}
    \label{fig:image1}
\end{figure*}

\begin{figure*}
    \gridline{\fig{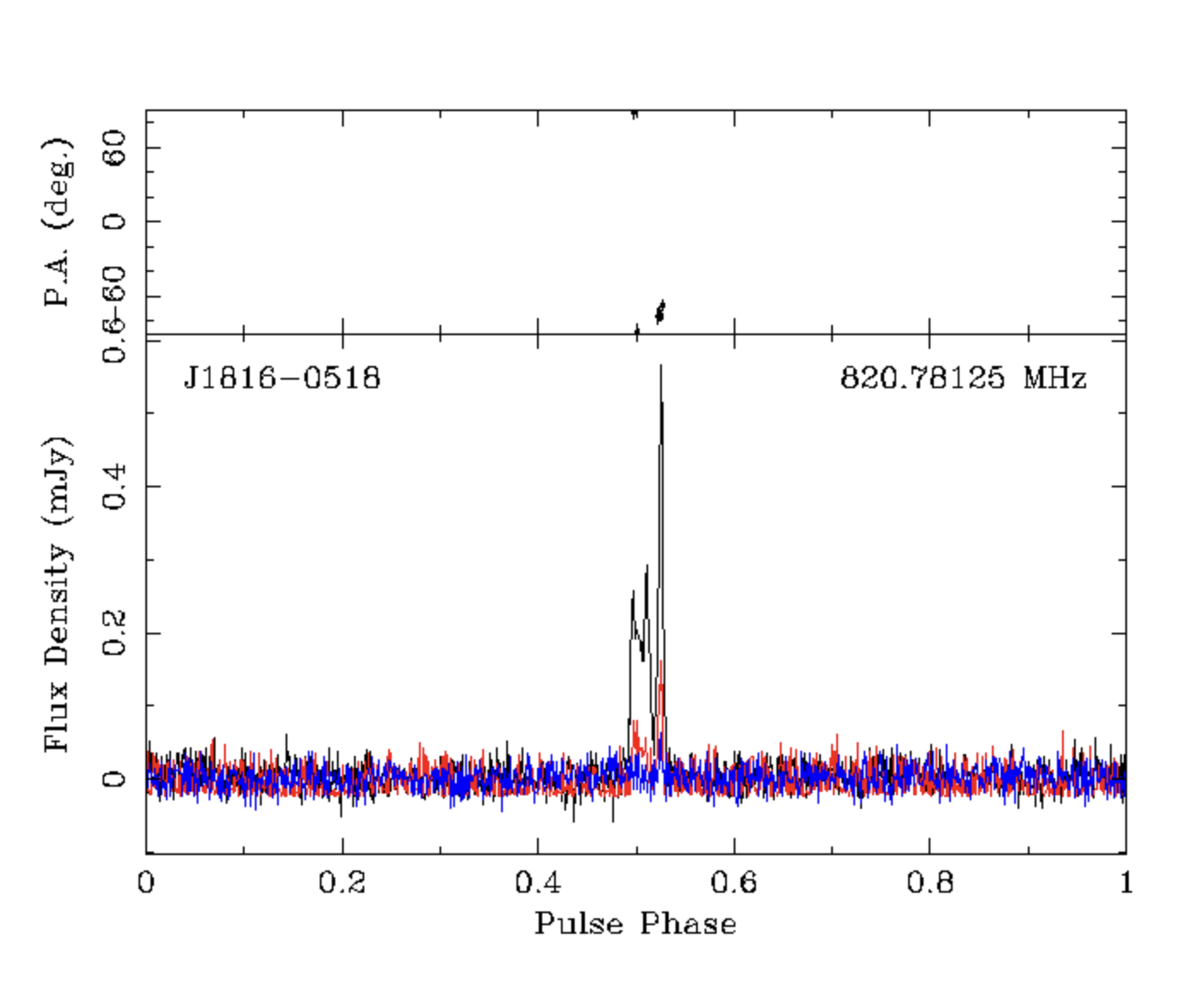}{0.49\textwidth}{(a) GBT 820-MHz composite profile based on  4.44 hours of data.}
             \fig{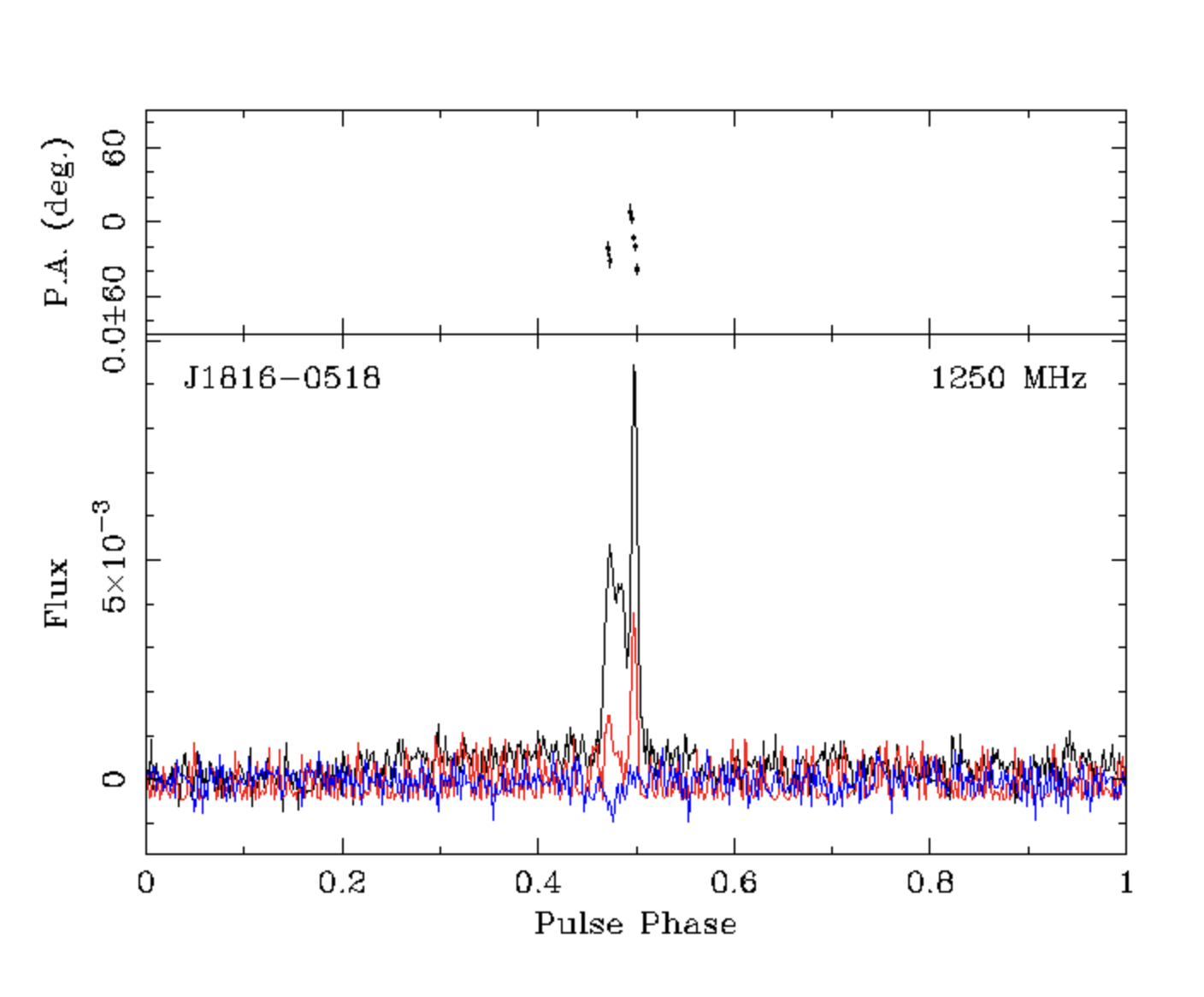}{0.49\textwidth}{(b) FAST L-band composite profile (in arbitrary flux units) based on 0.12 hours of data.}
            }
 \caption{Polarization and flux-calibrated composite profiles of PSR~J1816--0518 at different observing frequencies. (a) shows the  GBT profile at 820 MHz, and (b) is the FAST profile at L band (1250 MHz). Stokes parameters I, L, and V are represented by black, red, and blue curves, respectively.}
    \label{fig:image2}
\end{figure*}

\begin{figure*}
    \gridline{\fig{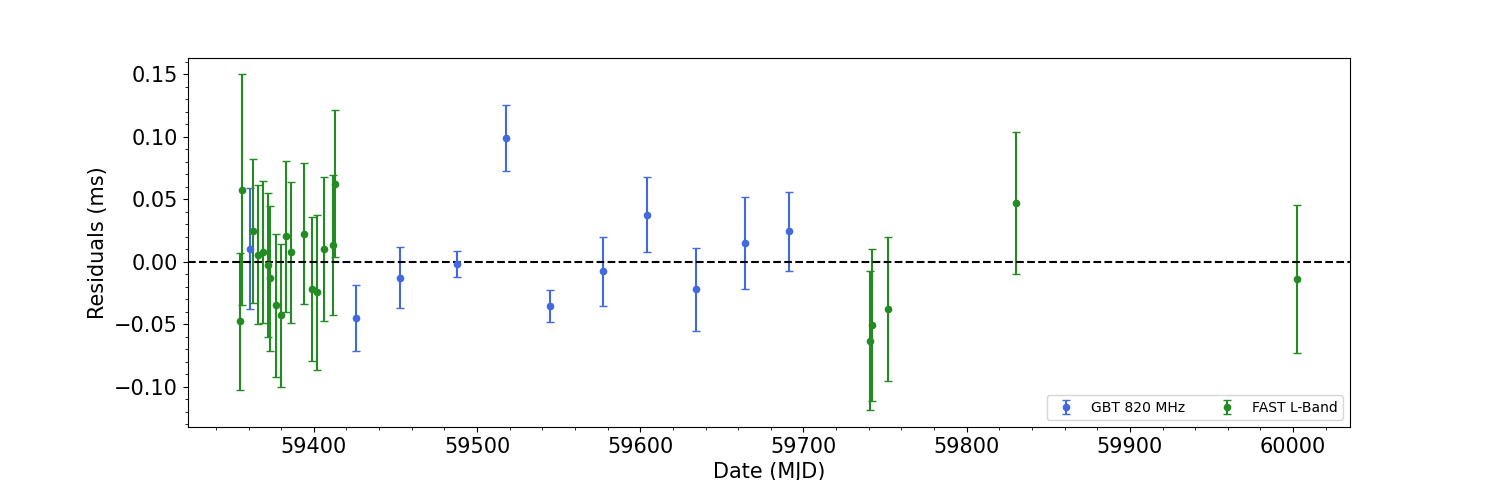}{0.99\textwidth}
     {(a)  PSR J0535--0231}
            }
    
    \gridline{\fig{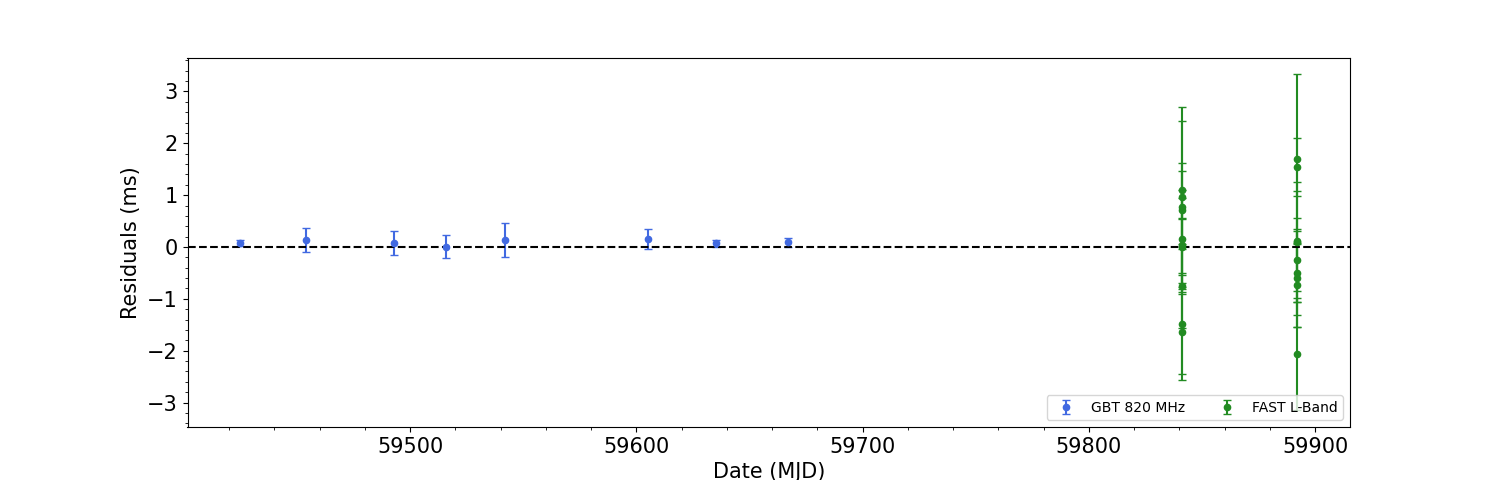}{0.99\textwidth}
    {(b)  PSR J1816--0518}
            }
    \caption{Post-fit residuals for (a) PSR~J0535--0231 and (b) PSR~J1816--0518. Across all observations for both pulsars, the GBT data at 820 MHz and FAST data at 1250 MHz are indicated by blue and green  points, respectively.}
    \label{fig:image3}
\end{figure*}


\section{Discussion \& Conclusions}

The position angle (PA) subplot in the FAST profile for PSR~J0535--0231  shown in Figure~\ref{fig:image1}(b), though not perfectly identical to the traditional S-shaped PA swing, has a central sloping region that appears to be approximately S-shaped. Therefore, we attempted to fit this feature to the Rotating Vector Model (RVM) \citep{RC_1969}, using the \texttt{PSRCHIVE} package, \texttt{psrmodel}, aiming to model the line-of-sight geometry of the pulsar beam. We first searched an 18x18 grid to constrain the parameters obtained from the fit -- namely, the viewing angle, $\zeta$, magnetic axis angle, $\alpha$, pulse phase, $\phi$ and position angle, $\psi$ -- and used the resulting best-fit values as starting parameters to re-fit to the model. This process resulted in a $\chi^2$ value well beyond 1, indicating that either the linear polarization sweep for this pulsar is not well-modeled by the RVM or that the signal-to-noise (S/N) is too low.

We calculated average flux densities for the 820 MHz GBT data by performing flux and polarization calibration for each pulsar. We obtained an estimated 820-MHz average flux density of 0.020 $\pm$ 0.003~mJy for PSR~J0535--0231 and 0.012 $\pm$ 0.002~mJy for PSR~J1816--0518. Flux calibration scans for GBT L-band and FAST L-band were not obtained.

In both pulsars, we find some noticeable profile evolution with frequency. For PSR~J0535--0231, the relative fluxes and linear polarizations of the trailing components increase with frequency. PSR~J1816--0518 shows rather minimal profile evolution, with its leading component appearing to have slightly higher linear polarization and relative flux than the trailing component at higher frequencies. The evolution of these aforementioned profile components typically arises from either intrinsic effects of the pulsar's magnetosphere, such as emission from higher altitudes at higher frequencies \citep{Pilia_2016},
or extrinsically, from scattering due to the propagation of emission through the ISM  \citep{Geyer_2017, Jing_2025}. These profile changes are inconsistent with those expected due to scattering and are, therefore, very likely to be intrinsic to the pulsar.

In Table~\ref{tab:timing}, we list the distances \citep{Kaplan_2022} derived by the NE2001 \citep{Cordes_2002} and YMW16 \citep{Yao_2017} models. Neither accurately models the electron density along the  distance to PSR~J0535--0231, as both models place it outside of the Galaxy. We calculate a DM of 118.1 $\pm$ 0.9 \dmunits but find that the maximum line-of-sight DMs predicted by NE2001 and YMW16 in this direction are 78.2 \dmunits and 117.1 \dmunits, respectively.
Our measured DM is nearly consistent with the YMW16 model, but not at all consistent with the NE2001 model. Both models ultimately overestimate the distance to this pulsar, returning the maximum model distances of 25 kpc and 44 kpc, respectively. This result is in agreement with previous studies showing that at similar latitudes, electron densities along the line of sight are larger for YWM16 than they are for NE2001 \citep{Deller_2019,Price_2021}.

Finally, the derived parameters obtained through pulsar timing, particularly spin period and period derivative, allow estimation of three characteristic parameters based on various relationships between $P$ and $\dot P$. The spin-down luminosity, $\dot E$, describes the rate of energy loss due to magnetic dipole radiation, the inferred surface magnetic field, $B$, describes the strength of the pulsar's magnetic field assuming an orthogonal rotator and the characteristic age, $\tau$, 
defines the age of a pulsar under the assumptions that the sole mechanism for its spin-down is magnetic dipole radiation and that the pulsar's initial period is significantly less than its current period. 

With the aforementioned parameters, we can classify pulsars among the general population in a $P$--$\dot P$ diagram, as seen in Figure~\ref{fig:image4}, where the sloping diagonal lines describe $\dot E$, $B$, and $\tau$. Here, the current pulsar/neutron star population is distinguished by the type of source under consideration (i.e. radio pulsars (both isolated and binary), x-ray and $\gamma$-ray sources, neutron stars associated with  supernova remnants (SNRs), magnetars, RRATs, etc.). We note the location of FAST-discovered pulsars explored in this follow-up effort (including the aforementioned DNS and BW pulsars) indicated by the four stars located at various positions in the diagram among the general pulsar population, including other pulsars discovered in the CRAFTS Survey.

Of particular interest is the location of PSR~J0535--0231 on the $P$--$\dot P$ diagram. 
We find a $\dot P$ below  $10^{-17}$, fairly unusual for  pulsars of similar spin periods.  This implies a magnetic field strength of 5 $\times 10^{10}$ G and a characteristic age of over 1 Gyr. These parameters suggest the pulsar may have been partially recycled, a result of angular momentum transfer from a companion star during its red giant phase. However, given that this pulsar appears to be isolated, it is likely that the companion star's supernova disrupted the binary, forming a disrupted recycled pulsar (DRP, \citealt{Lorimer_2_2004, Fiore_2023}). Using the ATNF Pulsar Catalog\footnote{\url{http://www.atnf.csiro.au/research/pulsar/psrcat}} \citep{Manchester_2005}, we estimate that DRPs have periods ranging between approximately 20 ms and 600 ms, along with rough $\dot P$ estimates between 2$\times$10$^{-20}$ and 2$\times$10$^{-17}$. Within this range restriction we find evidence, based on comparative histograms of the transverse velocities and distances for DRPs and DNS binaries, to support the claim previously made in \cite{Belczynski_2010} that DRPs tend to have higher transverse velocities than DNS binary pulsars. In this dataset, we find that median transverse velocities for the DRPs and DNS binaries are approximately 70 km s$^{-1}$ and 27 km s$^{-1}$, respectively. With higher kick velocities, DRPs should, then, be located at greater distances, on average, from the Galactic plane than their binary counterparts. By extension, the median distances of the DRPs and DNS binaries within the aforementioned range are 1.54 kpc and 1.1 kpc, respectively. Therefore, under the assumption that PSR J0535--0231 is a DRP, this may be a plausible explanation for its derived DM distance reaching the maximum values of both the NE2001 and YMW16 models in this direction of the Galaxy.

To date, FAST has discovered over 900 pulsars since first light in 2016, with nearly 80 of approximately 222 CRAFTS pulsars currently published \citep{FAST-CRAFTS}.
Of these 80 published pulsars, approximately 64 of them have timing solutions achieved through FAST \citep{Zhang_19,Wen_22,Miao_2023,Wu_23_1,Chen_23,Wu_23_2,Zhao_24} Arecibo \citep{Miao_22}, Parkes \citep{Cameron_20}, and Effelsberg \citep{Cruces_21} follow-ups. Given the vast number of pulsars detected in the FAST CRAFTS survey alone, it will likely take several years of follow-up efforts to produce initial timing solutions. With the growing number of new pulsar discoveries, it is likely that some of these may be faint, distant pulsars, like PSR~J0535--0231, that push the line-of-sight DM boundaries of the NE2001 and YMW16 electron density models beyond their current limits, giving rise to significant improvements in electron density modeling––critical both for pulsar and fast radio burst science.

In summary, we present two such pulsars first discovered by FAST and further observed using the GBT. We calculated TOAs and fit initial timing solutions spanning approximately 1.8 years total for PSR~J0535--0231 and just under 1.5 years for PSR~J1816--0518, making use of initial FAST follow-up observations in addition to the GBT campaign on which these results are primarily based. Flux- and polarization-calibrated profiles for the GBT 820-MHz observations were summed across epochs to produce composite, average profiles spanning the duration in which the pulsar was observed during the campaign. Composite profiles for the GBT L-Band observation for J0535--0231 and FAST observations of both pulsars are also included. Moreover, both pulsars exhibit some frequency-dependent profile evolution that could be further explored with additional multi-frequency observations.   

Distance estimates, based on the DM of J0535-0231, using the NE2001 and YMW16 electron-density models are not well-determined due to the pulsar's high latitude which yields overestimates from both models–– the degree of discrepancy depending on the specific model used. Ideally, more distant, faint pulsars will encourage improvements in this model in high-$b$ regions of the Galaxy, in addition to the other regions mentioned in \cite{Price_2021}.

Finally, 
we present estimates of each pulsar's magnetic field strength, spin down luminosity and characteristic age determined by measured values of $P$ and $\dot P$. Based on these values, we then place each FAST-discovered pulsar among the general pulsar population and
find that PSR J0535--0231 may be a partially-recycled pulsar, likely a DRP, given its isolated nature and timing-derived measurements of inferred surface dipole magnetic field and characteristic age.

Given the 200+ FAST CRAFTS pulsars and additional 900+ FAST-discovered pulsars, in general, we expect that there will be numerous future long-term follow-up efforts to obtain initial timing solutions for many sources as well as opportunities to include new FAST-discovered pulsars in improved DM distance modeling and carry out more in-depth analysis of special features of specific pulsars, such as beam geometry, profile evolution and scintillation. 

\begin{figure*}[!htb]
    \centering
    \includegraphics[width=0.9\linewidth]{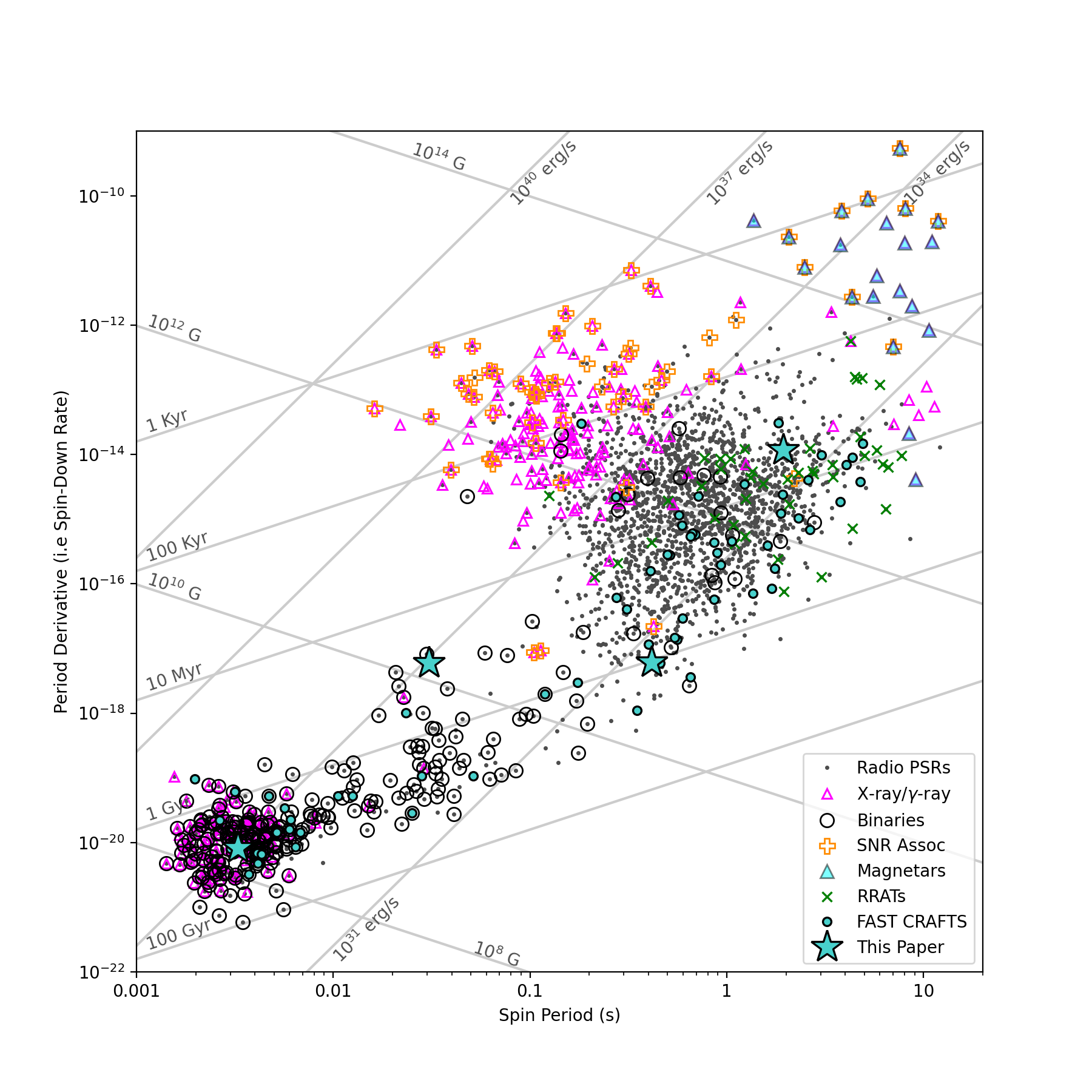}
    \caption{$P - \dot P$ diagram where FAST-discovered pulsars are placed among the general pulsar population. This figure was produced using the \texttt{PRESTO} code, \texttt{ppdot\_plane\_plot.py}\textsuperscript{2}. FAST CRAFTS pulsar (teal circles) parameters were obtained using the ATNF Pulsar Catalog.}
    \label{fig:image4}  
\end{figure*}

\newpage

\section*{Acknowledgments}
This work is partially supported by NSFC grant No. 12588202. V.A.B. acknowledges support from the Du Bois and Chancellor's Scholar Fellowship programs at West Virginia University. V.A.B. and M.A.M. are members of the NANOGrav Physics Frontiers Center (\#2020265). D.L. acknowledges the support from the New Cornerstone Science Foundation.
The National Radio Astronomy Observatory and Green Bank Observatory are facilities of the U.S. National Science Foundation operated under cooperative agreement by Associated Universities, Inc. This work has made use of the Harvard Astrophysics Data System, (\url{https://ui.adsabs.harvard.edu/}), the Australian Telescope National Facility (ATNF) pulsar catalog (\url{http://www.atnf.csiro.au/research/pulsar/psrcat}) and the Pulsar Survey Scraper (PSS) DM Model function (\url{https://pulsar.cgca-hub.org/compute}) as well as pulsar data reduction software: \texttt{TEMPO2} (\url{https://www.atnf.csiro.au/research/pulsar/tempo2}), \texttt{PRESTO} (\url{https://github.com/scottransom/presto}) and \texttt{PSRCHIVE} (\url{https://psrchive.sourceforge.net}).

\pagebreak
\pagebreak
\bibliography{references}
\bibliographystyle{aasjournal}

\end{document}